\documentclass[12pt,preprint]{aastex}

\shorttitle{Optical and NIR Structure Galaxies at z$\sim$0.3}
\shortauthors{La Barbera et al.}

\received{2001 December 3}
\begin{document}

\title{Optical and Near-Infrared Structural Properties of Cluster Galaxies at
 $z\sim$0.3\footnote{Based on observations collected at European Southern Observatory
(ESO n. 62.O-0369, 63.O-0257, 64.O-0236) and on data from the STScI Science Archive}}

\author{F. La Barbera}
\affil{Physics Department, Universit\`a Federico II,
    Napoli, Italy}

\email{labarber@na.astro.it}

\author{G. Busarello, P. Merluzzi and M. Massarotti}
\affil{Osservatorio Astronomico di Capodimonte, Napoli, Italy}

\email{gianni@na.astro.it}
\email{merluzzi@na.astro.it}
\email{michele@na.astro.it}

\and

\author{M. Capaccioli}
\affil{Physics Department, Universit\`a Federico II, Napoli, Italy}
\affil{Osservatorio Astronomico di Capodimonte, Napoli, Italy}
\email{capaccioli@na.astro.it}

\begin{abstract}

Structural parameters (half-light radius $r_e$, mean effective surface
brightness $\langle\mu\rangle_e$, and Sersic index $n$, parameterizing the light profile 
shape) are derived for a sample of galaxies in the rich cluster AC\,118 at 
$z=0.31$: so far the largest (N=93) sample of galaxies at intermediate-redshift 
with structural parameters measured in the near-infrared.
The parameters are obtained in two optical wavebands ($V$ and $R$) and in
the $K-$band, corresponding approximately to $B$, $V$ and $H$ rest-frame.
 
The distributions of $r_e$ at $z=0.31$ match those for the Coma cluster
(i.e. for the local universe) both in the optical and in the NIR.
The $K-$band distribution is of particular interest, since the NIR light mimics 
the mass distribution of galaxies.
The similarity of the distributions for the two clusters (AC\,118 and Coma)
proves that the galaxies at the bright end of the luminosity function did not 
significantly change their sizes since $z\sim0.3$ to the present epoch.

The ratio of the optical to the NIR half-light radius shows a marked trend 
with the shape of the light profile (Sersic index $n$). In galaxies with 
$n\gtrsim4$ (typical bright ellipticals) $r_{e,NIR}\sim 0.6 r_{e,opt}$, while 
the average ratio is 0.8 for galaxies with lower $n$ (typical disk systems). 
Moreover, the NIR Sersic index  is systematically larger than in the optical 
for $n\lesssim4$. These results, translated into optical and optical-NIR color 
gradients, imply that the optical color gradients at $z\sim$0.3 are similar 
to those of nearby
galaxies. The optical-NIR color gradients are in the average larger, ranging from -0.73 
mag/dex for $n\lesssim4$ to -0.35 mag/dex for $n\gtrsim4$.

Models with `pure age' or `pure metallicity' gradients are unable to reconcile
our color gradients estimates with observations at $z\sim 0$, but we argue that the 
combined effects of age and metallicity might explain consistently the 
observed data: passive evolution (plus the possible effect 
of dust absorption) may account for the differences between the optical and 
NIR structural properties.

The lack of any major change in $r_{e,NIR}$ since $z\sim0.3$ suggests that
merging involving bright galaxies did not play a significant role in the last 
$\sim$4.4 Gyr 
($\Omega_M$=0.3, $\Omega_{\Lambda}=0$, H$_0=50$ km s$^{-1}$Mpc$^{-1}$).

The results of the present paper will be applied to the study of the
scaling laws in subsequent works.

\end{abstract}


\section{Introduction}

Our knowledge on the evolution of galaxies comes from the study of the
properties of families of galaxies at different redshifts.  Observable
quantities like magnitudes, colors, structural parameters, velocity
dispersions and line strengths are combined into well-established
relations like, for instance, the color-magnitude relation
(e.g. Bower, Lucey, \& Ellis 1992; Kodama et al. 1998), the
Mg-$\sigma$ relation \citep{ZIB97}, the fundamental plane (Dressler et
al. 1987; Djorgovski \& Davies 1987; J\o rgensen, Franx, \& Kj\ae
rgaard 1996) and its projection in the $r_e$ -- $\langle\mu\rangle_e$
plane, that is the Kormendy relation (Kormendy 1977; Capaccioli, Caon,
D'Onofrio 1992; Ziegler et al. 1999). These relations can now be
studied over a significant fraction of the age of the universe.
Studies of the properties of early-type galaxies in rich clusters lead
to a large amount of observational evidence in favor of the passive
evolution scenario, in which an early ($z>$2) and short ($\sim$1 Gyr)
burst of star formation is followed by a long period where the aging
of stellar populations drives the observed properties of galaxies
through intermediate redshifts until the present epoch (Kodama \&
Arimoto 1997; Stanford, Eisenhardt, \& Dickinson 1998; J\o rgensen et
al. 1999; Kelson et al. 2000a, and references therein).  On the other
side, the alternative scenario of the hierarchical merging seems also
consistent with most of the existing data (e.g. Kauffmann \& Charlot
1998), indicating either that the present observational evidence is
not sufficient to discriminate between the two pictures or that galaxy
evolution is driven by both mechanisms.

The half-light radius and the mean surface brightness of a galaxy
measure the size and the density of the luminous matter.
These quantities are related both to the gravitational potential of the
galaxy and to the properties of the stellar populations.
The relations existing between the structural parameters of galaxies have 
been recently studied at intermediate redshifts, both 
in the form of $r_e$ vs. $\langle\mu\rangle_e$
\citep[and references therein]{ZSB99} and as luminosity-size relation  
\citep{SLL96,SKF99}. 
\citet{SLL96} found that disks at $z>$0.5 are in the average 
$\sim$ 1.5 mag brighter than in local galaxies.
This was contrasted by the results of \citet{SKF99}
who found no luminosity evolution in disks up to $z\sim$0.9.
\citet{ZSB99} found that the Kormendy relation of cluster galaxies
at $z=0.4$ and $z=0.55$  is consistent with the passive evolution scenario.
The Kormendy relation has also proven to be a powerful tool to test the 
expansion of the universe by the measurement of the cosmological
dimming (i.e. for the Tolman test). \citet{LUB01} studied the
Kormendy relation up to $z=0.92$, finding that the data are consistent with
the cosmic expansion together with passive evolution.
Finally, the structural parameters enter the fundamental plane of
early-type galaxies, of which the Kormendy relation is a
two-dimensional projection.
Due to its very small scatter, the fundamental plane is a powerful 
tool to study the evolution of the M/L ratio of galaxies belonging 
to distant clusters. The recent determinations of the fundamental plane
at intermediate redshifts constrain the epoch of star formation 
in early-type galaxies at large redshifts ($z>$2 for $\Omega_M$=0.3 or 
$z>$5 for $\Omega_M$=1 and $\Omega_{\Lambda}=0$; 
see van Dokkum \& Franx 1996; J\o rgensen et al. 1999; Kelson et al. 2000a).
However, these studies have to face with the problem of the 
progenitor bias, i.e. with our ignorance of the relationship between local
galaxies and their high redshift counterparts (see van Dokkum \& Franx 2001
for a thorough discussion).

A clue for the interpretation and for the applications of the scaling laws 
is the dependence of the structural parameters on wavelength.
The basic distinction between optical and NIR is that the 
optical wavebands are very sensitive to the effects of metallicity 
(through line-blanketing) and to the content of younger populations 
(dust absorption included), while the NIR wavebands are
dominated by the old, more quiescent, stellar populations and 
are virtually unaffected by dust absorption.
For these reasons the NIR is the best 
tracer of the mass distribution inside a galaxy. 
Local galaxies are smaller at NIR wavebands than 
in the visible, with a similar, though weaker, trend also within the optical
wavelength range (Scodeggio et al. 1998; Pahre, de Carvalho, \& Djorgovski 
1998).
This behavior is confirmed by the relatively large 
optical-NIR color gradients measured 
in early-type galaxies, the optical-optical
gradients being usually much smaller, except when the $U-$band is involved
(Peletier et al. 1990a; Peletier, Valentijn, \& Jameson 1990b; 
Hinkley $\&$ Im 2001; Nelson et al. 2001).
Small optical color gradients were recently measured in distant 
(up to $z\sim$1) cluster early-type galaxies by \citet{SMG00} and by  
\citet{TKA00}. These authors concluded that metallicity is the primary
factor driving the color gradients.

The derivation of the structural parameters at intermediate redshifts
is affected by the problem of the small angular size of the galaxies,
whose half-light radii are typically less than 1 arcsec.
For this reason many works were based on imaging taken with the HST.
It has been demonstrated, however, that reliable structural parameters
can be derived from ground-based photometry as well, on high
resolution ($< 0.1 ''$/pxl) and exceptional seeing ($<0.4''$)
conditions, provided that the properties of the PSF are carefully
taken into account \citep{SLL96,JOR99}.

In this paper we obtain the first multi-band large set of structural
parameters for galaxies at an intermediate redshift ($z=0.31$).  The
data are derived from ground-based (ESO-NTT) photometry in two optical
($V$ and $R$) and one NIR ($K$) wavebands.
The derived values for the structural parameters will be published in a
forthcoming paper addressed to the study of galaxy evolution by
means of the scaling laws (Busarello et al. 2001a, in preparation).

The paper is organized as follows.
The sample and the data used in this study are presented in 
\S~\ref{samples}.
The derivation of the structural parameters is outlined in 
\S~\ref{STRSEC}. The structural parameters from HST imaging are derived in 
\S~\ref{HST_HST}, where the results from Sersic models and bulge-disk decomposition
are compared.
In \S~\ref{SEC_HST_R} we use HST photometry for a subset
of galaxies to verify the reliability of ground-based structural parameters.
The structural parameters are presented in \S~\ref{OP_NIR}, where we make
a comparison between optical and NIR properties at $z\sim0.3$ and with the values 
at $z\sim0$. In \S~\ref{SECGRAD} we analyze the optical
and NIR structure of galaxies in terms of internal color gradients.   
The results are discussed in \S~\ref{CONC}.

\section{Description of the Sample} \label{samples}

AC\,118 (Abell 2744) is a rich (`Coma-like') galaxy cluster at redshift $z=0.31$
that was studied by several authors in relation to its galaxy
population, X-ray and radio properties and gravitational lensing
\citep{BAR96,CBS98,SED98,WNB99,WXF99,ALL00,GEF01}.

New photometric data and medium resolution spectroscopy has been collected
during three observing runs (October 1998 - September 2000) at the ESO
New Technology Telescope.
The data of interest for the present work are the imaging in the $V$
and $R$ wavebands obtained with the EMMI instrument and the imaging in
the $K$ waveband obtained with SOFI.  The photometry is complemented
with HST archive data consisting of a single pointing on the cluster 
center. 
The relevant information on the data is given in Table~\ref{AC118_DATA}.
The reduction of the $K-$band image is described in \citet{AND01}, while
the reduction of the NTT optical data will be presented elsewhere 
(Busarello et al. 2002b, in preparation). 
The HST data were retrieved from the STScI Archive and reduced following
the steps described in \citet{CBS98}. The NTT  photometry
was used to calibrate the HST imaging into the Johnson $R-$band as 
outlined in Appendix A. 

The samples used in the present study consist in: A) 50 
galaxies in the HST field; B) 37 galaxies in the HST field in common with 
the ground-based $R-$band sample; C) 58 galaxies in common in the $V-$ and 
$R-$band; D) 93 galaxies in common in the $R-$ and $K-$band.

Samples C and D are the results of two selection steps.
The first selection was based on photometric redshifts \citep{MIB01},
as derived from the above photometry plus $I-$ and $U-$band (new NTT data
and NTT archive data respectively). The reliability of the 
photometric redshifts was verified on 75 galaxies 
with available spectra (Merluzzi et al. 2002, in preparation).
Morphology is not a selection criterion in this study, since a morphological
classification is not possible from our data.
The photometric redshifts algorithm flags the galaxies according to
whether their colors are consistent with early- or late-type template
spectra.
We did not use this classification as a selection criterion, and retained in the
samples all the galaxies with the correct photometric redshift.
We remark, however, that only 7 galaxies out of 93 in sample D
are `late-type' according to their photometric properties (from $U$ to $K$).
The second selection step is based on the reliability of the measurement 
of the structural parameters and is the result of the analysis presented
in \S~\ref{SIMSEC}. This lead to include in the sample the galaxies 
brighter 
than $R$=21 and $V$=20.5, and to exclude objects with an estimated $r_e$ 
smaller than 0.15 arcsec. 

Samples A and B were selected in order to obtain 1) reliable structural 
parameters in the NTT $R-$band and 2) the largest number of galaxies in the 
HST field with the same range of sizes and luminosities of the ground-based 
samples.
This lead to include galaxies brighter than $R$=21 and to exclude objects 
with $r_e$ smaller than 0.15 arcsec both in the NTT and in the HST image.
A further (a posteriori) selection was done for all the samples
on the basis of the results of the fitting: 
objects with very large uncertainties ($>$100\%) were not considered in the
analysis. This however lead to exclude only 2\% of the objects.   

Morphological classification from HST imaging by \citet{CBS98}
is available for 21 out of 93 galaxies in sample D (17
out of 58 in sample C).
Out of these, 6 (6) are classified as ellipticals, 10 (7) as S0, and 5 (4) as 
spirals. These numbers reflect the overall distribution of morphological
types in AC\,118, which is dominated by early-type (mainly S0) galaxies
\citep{CBS98}.
The question of morphological types will be resumed in the conclusions.
The relevant information on the samples is summarized in Table 2.

\section{Derivation of the structural parameters} \label{STRSEC}

Since the typical galaxy sizes at $z\sim$0.3 are smaller than $\sim
1''$, it is crucial to take into account the effects of the Point
Spread Function (PSF) when deriving structural parameters from ground-based data. 
This means that a proper model for the galaxy surface brightness 
(SB) has to be convolved with the PSF and fitted to the observed SB. 
This is usually performed: 1) by one-dimensional fitting algorithms
(e.g. Saglia et al. 1997) or 2) by two-dimensional methods 
\citep{VDF96,SKF99}.
In case 1) the integrated light curve or the SB profile
are derived by an elliptical fit to the galaxy isophotes and then fitted
with the adopted model.
In the two-dimensional approach, a direct fit of the galaxy SB is performed. 
Advantages and drawbacks of both methods have been largely discussed
in previous works (see Kelson et al. 2000b, and references
therein). In the present study we adopt the two-dimensional fitting
approach for two reasons: 1) many galaxies in
AC\,118 belong to overlapping systems, and only the 2D approach can deal
properly with these cases; 2) isophotal fitting 
of the ground-based images should be performed on a very small number of 
pixels, resulting in poorly constrained isophotes. 

The SB of distant galaxies is well described 
by two-dimensional models with constant ellipticity and position angle. 
In this study, we use two kinds of models: the Sersic model (Sersic 1968)  
and the ($r^{1/4}-$) bulge + disk ($B+D$) model. Since the $B+D$ 
decomposition requires high resolution and signal to noise ratios  
(see Ratnatunga, Griffiths, \& Ostrander 1999), this model is applied only 
to the HST data. Ground-based 
structural parameters are derived by fitting the Sersic models. 
The models used for the PSF of the NTT and HST images are presented in 
\S~\ref{PSFSEC} while the fitting algorithm and the galaxy models 
are described in \S~\ref{FITSEC}.
The limits of applicability of the fitting algorithm are set in
\S~\ref{SIMSEC}.

\subsection{The Point Spread Function} \label{PSFSEC}

To model the PSF of the HST image, we used the Tiny Tim 5.0 software
package \citep{KRI99}.  Since the PSF depends on the position on the
field, different models were created for each galaxy at the
corresponding chip location.
Effects of PSF under-sampling and broadening from image jittering are
smaller than few percents of $r_e$ for the typical sizes of galaxies
at this redshift and therefore these were ignored (see Kelson et al. 2000b). 

To model the PSF of the NTT images we proceeded as follows. Stars in the 
cluster field were selected by the stellar index parameter $SG$ of S-Extractor 
\citep{BER96} and from the aperture magnitude. 
To minimize star -- galaxy classification errors, we considered only objects 
with $SG \ge 0.99$. We selected sources with $K \le 18$ mag in the SOFI
field and $R \le 21$ mag in the region not covered from the infrared imaging. 
Such criteria allowed to obtain a sufficient $S/N$ to model the PSF.
The reliability of the selection is also supported by the photometric redshift 
algorithm: the colors of the candidate stars do not match galaxy 
template spectra at any redshift. 

The PSF of the NTT images is well described by a sum of two Gaussians 
convolved with the pixel function (see appendix~\ref{app_gpsf} for details). 
The model has nine relevant parameters: the center coordinates, the
central intensity, the Full Width at Half Maximum (FWHM) of the two 
Gaussians in the two
perpendicular directions, the ratio of the central intensities, and
the local value of the background.
The fits were performed by a $\chi^2$ minimizing algorithm based on
the Levenberg-Marquardt method \citep{PFT86}.
Typically, the residuals from
the fitting are smaller than  5\%. The uncertainties on fitting 
parameters were estimated by numerical simulations on models with Gaussian 
noise. For each pixel, the standard deviation of the noise was computed 
by the distribution of the fit residuals in the adjacent pixels\footnote{
This allows to take into account empirically the various sources of 
uncertainties: flat-field accuracy, photon and read-out noise, 
deviations from the fit model, etc. \label{SN} }.
Figure~\ref{PSFFIT} shows the values of the PSF parameters in the
$V-$, $R-$ and $K-$band as a function of the distance to the image centers.
In general the PSF tends to become broader as this distance increases.
The effect is very weak in the $K-$band image, while significant
variations are found in the optical images.
The trends of the parameters were modeled with empirical analytic curves, 
that were then used to obtain the PSF model relative to each galaxy position.

\subsection{Modeling of the brightness distribution} \label{FITSEC}

Galaxy images were fitted by the following model:
\begin{equation}
M(BG, \left\{ p_k \right\} ) = BG + B(\left\{ p_k \right\}) \, \circ S \ \ ,
\label{GALMOD}
\end{equation}
where $B$ is the galaxy brightness distribution, that is described by a 
set of parameters $\left\{ p_k \right\}$, $S$ is the PSF model, $BG$ is 
the value of the local background, and the symbol $\circ$ denotes the 
convolution. 
For the Sersic models, the parameters $\left\{p_k \right\}$ are: the 
effective major 
semi-axis $R_e$, the central surface brightness $I_0$, the Sersic 
index $n$, the axial ratio $b/a$, the position angle $PA$, the coordinates 
of the photometric center and the local value of the background. 
In the case of the $B+D$ decomposition, the parameters that enter the fit are 
those of two Sersic models with $n=4$ and $n=1$ for the bulge 
and the disk respectively.
The fit parameters are used to derive the total magnitude $m_T$ (e.g. 
Saglia et al. 2000), $r_e$ and $\langle\mu\rangle_e$. For Sersic models, $r_e$ is 
computed as the equivalent radius of the ellipse that encloses half of the 
total galaxy light, i.e. $r_e=\sqrt{b/a} \cdot R_e$. For the $B+D$ 
decomposition, $r_e$ is calculated from the growth curve of circular 
aperture magnitudes. In both cases, we have 
$\langle\mu\rangle_e = 2.5 \log(2\pi) + 5 \log(r_e) + m_T$.
For the Sersic models we found that both methods of computing $r_e$ 
agree within $2 \%$. Therefore, the difference of the two definitions can 
be ignored in the following analysis. 

The parameters $\left\{ p_k \right\}$ and $BG$ were derived by a least squares 
fit, by minimizing the function
\begin{equation}
 \chi^2( BG, \left\{ p_k \right\} ) =
 f( \left\{ p_k \right\} )  \cdot \left[
  \sum_j \left( I_j  - M_j(BG, \left\{ p_k \right\} ) \right)^2 
   \right] \ , 
\label{CHIEQ}
\end{equation}
where $I_{j}$ is the observed brightness distribution, $j$ is an index
running on image pixels and $f$ is a penalty function introduced to
avoid `not allowed' regions of the parameter space (i.e. unrealistic
values that can be found on local minima of $\chi^2$).
The $\chi^2$ minimization was performed by the Levenberg-Marquardt algorithm. 
The convolution with the PSF was computed straightly in the real domain in order to 
avoid high frequencies problems when convolving galaxy models 
by the Fast Fourier Transform.
Bad pixels, degraded regions and pixels contaminated by background
objects, were excluded from summation in equation~\ref{CHIEQ}. To this
aim, a masking image was created interactively for each galaxy.  
Overlapping objects were fitted simultaneously. 

Uncertainties on structural parameters are usually estimated either from the
formal errors on fitting results or by numerical simulations.
The formal uncertainties are derived by the topology of the $\chi^2$ function 
near to the solution and therefore they do not take into account the 
various sources of errors on the estimated parameters (e.g. Smith et al. 
1997).  Numerical simulations give more reliable uncertainties but they 
depend on the adopted galaxy model and are much more expensive in terms 
of computation time. 
As a compromise between accuracy and amount of computation time, 
we used a first order approximation: the uncertainties on the fit parameters 
were estimated from the derivatives of the $\chi^2$ function taking into 
account the noise in the galaxy image and the uncertainties on the PSF 
parameters. Details are given in Appendix~\ref{app_err}.

\subsection{Simulations} \label{SIMSEC}

To test the performance of the fitting procedure and to set the limits 
to its applicability, we created two sets of simulations based on Sersic 
$r^{1/n}$ models convolved with the NTT $R-$band PSF and with the typical 
noise of ground-based data. To take into account the uncertainties on the 
PSF, all the simulations were generated with the same PSF but the fits were 
performed with different PSF models whose parameters were generated in 
agreement with the corresponding uncertainties (see \S~\ref{PSFSEC}).  

The first set of simulations consist in $r^{1/4}$ galaxies
spanning the full magnitude range in the $R-$band (19 mag $\le m_T \le
23$ mag). Two values of $r_e$ were considered: $r_e = 0.3''$ and $r_e = 0.6''$, 
that correspond to the sizes of a small and a typical galaxy in a cluster 
at $z\sim 0.3$ (e.g. Kelson et al. 2000a). 
In Figure~\ref{SIM_MTOT} we give the results of the simulations in terms of 
differences between the input model parameters and the output from the fitting 
procedure. The quantities shown are $r_e$, the axis ratio $b/a$ (both given as 
logarithms), $m_T$, $\langle\mu\rangle_e$, $n$, and the `fundamental plane
parameter', $FP= \log r_e - 0.3 \times \langle\mu\rangle_e$, that is the combination of 
the structural parameters entering the fundamental plane.
It can be seen that the scatter of the fit parameters 
around the `true' values  
is relatively small for $m_T<$22 where it amounts at most to 25\% 
in $r_e$, 0.5 mag in $\langle\mu\rangle_e$, 0.15 mag in $m_T$, $15\%$ in $FP$ and
$10\%$ in $b/a$. The scatter of $n$ is higher, 
and varies from $ \Delta n \sim 1 $ at $m_T=19$ mag up to 
$ \Delta n \sim 3 $ at $m_T=22$ mag. 
Objects fainter than $m_T=22$ mag are affected by larger and systematic 
differences between input and computed parameters. 

The second set of simulations was created in order to test the fitting 
algorithm with respect to the galaxy size. To this aim, a wide range of values 
of $r_e$ was explored ($r_e=0.05''$ to $r_e=3''$) for three Sersic models
with $n = 1,4,7$ respectively. For each value of $r_e$, a suitable value of 
$\langle\mu\rangle_e$ was derived from the Kormendy relation, whose 
slope and the zero point were calibrated a posteriori 
from the present sample. 
In Figure~\ref{SIM_RE}, we show the comparison between true and estimated 
parameters. 
An inspection of Figure~\ref{SIM_RE} shows that the parameters are
recovered with a relatively small scatter (see above) and without
systematic errors for $r_e>0.15''$ ($\log r_e > -0.8$) for the whole
range of $n$ values.
The shape of light profile is subject to a small systematic error 
($\Delta n \sim 1$)
for $n=7$ below $r_e \sim 0.5''$ ($\log r_e \sim -0.3$). We also notice 
that the scatter increases for large $n$ values, due to the effect of the 
uncertainties on the PSF for galaxies with more concentrated profiles. 
 
The above results were used to derive the selection of 
the samples in this study.
All the samples include the galaxies with $r_e>0.15''$ in each band; 
in the $R-$band only galaxies with $R<$21 mag were included as a conservative 
limit; 
in the $V-$band a more restrictive selection in magnitude ($V<$20.5 mag) was
necessary due to the lower S/N, while no further restriction was
needed for the $K-$band photometry. 

\section{HST structural parameters} \label{HST_HST}

Structural parameters were derived for $N=50$ galaxies in the HST image
(sample A of Table~\ref{tab_SEL}), 42 of these objects are cluster members 
according to the photometric redshifts. 
Galaxies were fitted by both Sersic and $B+D$ models, and a flag
$F$ was assigned to each galaxy for the different cases that were encountered 
during the fitting.
For $N=5$ galaxies, both models gave small fitting residuals, but the $B+D$ 
decomposition produced meaningless values of the structural parameters for the 
bulge or for the disk component (e.g. $r_e < 0.02''$).  For such galaxies 
a possible second physical component is too weak or too small to be detected. 
However, except for one object that was excluded from the analysis, 
acceptable global parameters ($r_e$ and $\langle\mu\rangle_e$)
were obtained from the $B+D$ fit so that  
such  galaxies were retained in the sample and  were flagged with $F=-1$.
For $N=24$ objects, both Sersic and $B+D$ models gave small fitting residuals 
and acceptable estimates of the parameters. They were flagged with 
$F=0$.  $F=1$ was used to indicate $N=14$ galaxies for which residuals from the 
Sersic model shown evidence for a second component that was properly described
with the $B+D$ decomposition. The flags $F=0$ and $F=1$ were assigned by visual 
inspection of the galaxy fits as illustrated in Figure~\ref{BD_S}.
Finally, $N=7$ galaxies, of which  some objects with very distorted morphology 
and some BCGs with very large residuals ($>50\%$), were flagged with $F=-2$ and 
excluded from the present analysis. Fitting models suitable for such objects 
will be discussed in a forthcoming work. 

In Figure~\ref{BD_S_PAR}, we compare the structural parameters $r_e$, $\langle\mu\rangle_e$, 
$m_T$ and $FP$ obtained by different models. 
Table~\ref{tab_BD} shows the medians and the standard 
deviations of the differences in fitting parameters, and the relative bootstrap 
confidence intervals. 
By weighting each point in Figure~\ref{BD_S_PAR} with the fitting uncertainties
(\S~\ref{FITSEC}), we did not find significant variations of the estimates 
given in Table~\ref{tab_BD}.  This also holds for the other comparisons of 
structural parameters in the present work (see \S~4 and \S~5).

It is clear from Table~\ref{tab_BD} and from Figure~\ref{BD_S_PAR}
that the mean differences of the parameters have no significant
biases, also when galaxies with different flag $F$ are considered.  On
the other side, the scatter in the structural parameters depends on
the $F$ value: as expected, minimum values are obtained for $F=0$ when
$B+D$ and Sersic models give an equivalent description of the galaxy
image. The scatter varies from $35 \%$ to $ 50 \%$ for $r_e$, from
$\sim 0.8$ mag to $1.3$ mag for $\langle\mu\rangle_e$, and from $\sim
0.25$ mag to $\sim 0.43$ mag for $m_T$.  One important fact is the
robustness of the $FP$ parameter against the adopted fitting method:
due to the correlation between the errors in $r_e$ and in
$\langle\mu\rangle_e$ (Hamabe \& Kormendy 1987; J\o rgensen, Franx,\&
Kj\ae rgaard 1995), its scatter varies from $\sim 7 \%$ for $F=0$ to
$\sim 14 \%$ for $F=-1$.  The previous results are fully consistent
with the findings of Kelson et al. (2000a), although they find a
smaller dispersion ($\sim 3\%$) between the $FP$ parameters derived by
different fitting methods.  The results shown in Figure~\ref{BD_S_PAR}
and Table~\ref{tab_BD} give a measure of the intrinsic uncertainties
on the structural parameters of the galaxies in the present sample and
will be used as a reference in the following sections.

\section{Comparison of HST and NTT surface photometry} 
\label{SEC_HST_R}

The 2D fits for a representative subsample of galaxies in the HST and
NTT $R-$band images are shown in Figure~\ref{HST_R}.
Although the residuals from the fit are more evident in the HST images
due to the higher resolution and to the higher $S/N$, both HST and
ground-based images can be suitably fitted with the $r^{1/n}$ models,
also when overlapping objects are considered.

To test the reliability of the ground-based structural parameters, we
compared the estimates obtained by fitting Sersic models to the HST
and the NTT images of galaxies in the sample B of
Table~\ref{tab_SEL}. Since both filters cover approximately the same
spectral range (rest-frame $V-$band), we can rule out any difference
between the two datasets due to a possible waveband dependence of the
structural parameters.  
The structural parameters are compared in
Figure~\ref{conf_HST_R}, where we plot the same quantities as in
Figure~\ref{BD_S_PAR}. The medians and the standard deviations of the
differences in $\log r_e$, $\langle\mu\rangle_e$, $m_T$ and $FP$ are
given in Table~\ref{tab_HST_R}.  Table~\ref{tab_HST_R} and
Figure~\ref{conf_HST_R} prove that the mean differences of the
parameters have no significant biases.  The scatter between the two
sets of measurements is relatively small: it amounts to $\sim 38 \%$
for $r_e$, $\sim 0.7$ mag for $\langle\mu\rangle_e$, $\sim 0.3$ mag
for $m_T$ and $\sim 13\%$ for the $FP$ parameter.  It is worth to
stress again that the $FP$ parameter has significantly smaller
dispersion with respect to the other quantities.  The comparison with
the results of the previous sections proves that the differences in
the parameters derived from HST and from NTT photometry are of the
same order of those found when applying different fitting methods to
the HST image. In other words, the difference between the structural
parameters derived from HST and from NTT data is consistent with the
precision with which these quantities can be measured (see Kelson et
al. 2000a; J\o rgensen et al.  1999).  We thus conclude that reliable
structural parameters can be still derived for typical sizes and
luminosities of cluster galaxies at redshift $z\sim 0.3$ by using
ground-based data taken in ordinary observing conditions (i.e. seeing
$\sim1''$ and pixel scale $\sim 0.28 ''/$pxl).

\section{Optical and Near-Infrared Structural Parameters} 
\label{OP_NIR}

\subsection{Structure of Galaxies at $z\sim0.3$}

The $V-$ and $R-$band structural parameters 
for $N=58$ galaxies (sample C of Table~\ref{tab_SEL}) are compared in
Figure~\ref{V_R_PAR} and Table~\ref{tab_V_R}.
To allow a direct comparison, $\langle\mu\rangle_e$ and $m_T$
of $V-$band galaxies were corrected with the observed $V\!-\!R$ 
colors (Busarello et al. 2002b, in preparation).
$V-$ and $R-$band parameters show no significant systematic differences and
the scatter between the two sets of data is of the same order as those
found in the preceding sections, i.e. equal to the typical uncertainty of
structural parameter measurements.
The mean ratio $r_{e,R}/r_{e,V}\sim 0.97^{1.00}_{0.91}$ is consistent with
that found in local early-type galaxies ($r_{e,I}/r_{e,V}\sim 0.91$, 
Scodeggio 2001; see also Scodeggio et al. 1998). We defer further 
considerations to \S~\ref{SECGRAD}, where the results will be 
discussed in terms of color gradients.
Our $R-$band structural parameters can be directly compared to those derived at 
$z=0.33$ by \citet{KIB00} in the same wavelength range.
The full consistency of the two data sets is apparent from 
Figure~\ref{Phare} (upper panel) where the distributions of half-light radii 
are compared (the surface brightness is not shown for brevity reasons).

The $R-$ and $K-$band structural parameters for the 93 galaxies in
sample D (Table~\ref{tab_SEL}) are compared in Figure~\ref{R_K_PAR}
and Table~\ref{tab_R_K}. NIR $r_e$ and $\langle\mu\rangle_e$ are
significantly smaller than the relative optical parameters, while the
$n$ values are systematically higher.
On the average, such differences amount to -0.125 in $\log r_e$, to -0.5 mag 
in $\langle\mu\rangle_e$ and to +0.5 in $n$. 
A deeper inspection reveals the existence of 
two distinct subsamples which are separated by the value of the
Sersic parameter. 
For low $n$ values ($n< 4$), the differences of $K-$ to $R-$band
effective radii amount to -0.09 in $\log r_e$, and the differences in
the Sersic index are higher ($\Delta n \sim 1$ instead of 0.5).
On the other side, for $n > 4$ the Sersic indices are nearly the same in
the two wavebands ($\Delta n \sim 0$) while the $K-$band half-light
radii are much smaller than in the optical (-0.2 dex).
The confidence limits in Table~\ref{tab_R_K} show 
that the separation into the two classes is highly significant.
According to the Kolmogorov-Smirnov test, the probability that the two samples
come from the same distribution is less than 2\% for both $\Delta n$ and
$\Delta \log r_e$.

The results of the previous sections make us confident
that the systematic differences found between the structural
parameters are real and do not originate from biases induced by the
fitting method. No other measurements of $K-$band (rest-frame $H$)
structural parameters are available at $z\sim0.3$ so that no direct
comparison is possible. Again we defer further considerations 
to \S~\ref{SECGRAD}. 

\subsection{Comparison with Local Galaxies}

To make a comparison with the structural properties of nearby
galaxies, we made use of the sample of 70 galaxies belonging
of the Coma Cluster from the compilation by \citet{PAH99}. 
The distribution of the $V-$band $r_e$ from \citet{PAH99} is 
shown in Figure~\ref{Phare} (notice that our $R-$band corresponds to the
rest-frame $V-$band).  The radii of  \citet{PAH99} were scaled to $z=0.31$ 
by adopting $\Omega_M=0.31$, $\Omega_{\Lambda}=0$ (different 
cosmologies do not significantly alter the result).

The distributions of NIR $r_e$ are also compared in Figure~\ref{Phare} 
(the radii of Pahre 1999 were scaled as above). 
It is worth noticing that, although our rest-frame waveband
is different from the $K-$band in \citet{PAH99}, the two wavebands
actually probe the same stellar populations, so that the direct 
comparison has an immediate physical interpretation. 
The medians of the distributions in the NIR are $-0.22^{-0.18}_{-0.25}$ and 
$-0.22^{-0.18}_{-0.26}$ for Coma and AC\,118 respectively, while the 
corresponding values in the optical are $-0.12^{-0.08}_{-0.15}$ and 
$-0.08^{-0.04}_{-0.12}$.   
We thus conclude that the median of the half-light 
radii at $z\sim0.3$ is fully consistent with the median at $z\sim0$ 
both in the optical and in the NIR.

This result indicates that the sizes of bright galaxies did not
substantially change since $z\sim 0.3$ to the present epoch.  Since
the NIR probes directly the mass distribution inside galaxies, and
assuming that changes in mass produce corresponding changes in size,
we argue that during the last $4.4$ Gyr
\footnote{$\Omega_M$=0.3, $\Omega_{\Lambda}=0$, H$_0=50$ km
s$^{-1}$Mpc$^{-1}$} the bright cluster galaxies did not accrete a
significant mass fraction and that (dissipation-less) merging did not
play a major role (see also Nelson et al. 2001).

\section{Implications for the Optical and Near-Infrared 
Color Gradients} \label{SECGRAD}

The role played by the Sersic index in 
the relation between NIR and optical structural properties (\S~6.1) suggests
to discuss the combined behavior of half-light radii and Sersic indices
in terms of internal color gradients.

The available measurements of optical-NIR color gradients at intermediate
redshifts concern 6 field early-type galaxies studied in $V-H$ 
by \citet{HII01}. 
Unfortunately, no one of the galaxies is close enough in $z$ to our sample to
allow a significant comparison.
The comparison of recent measurements of optical color gradients at 
intermediate-redshifts with those in nearby galaxies shows 
little or no evolution since $z\sim1$, supporting the interpretation
in terms of metallicity gradients \citep{PVJ90,PDD90,SMG00,TAO00}.

We define the color gradient as $\Delta(\mu_2-\mu_1)/\Delta(\log r)$, where
$\mu_i$ is the surface brightness in the band $i$, and the differences are
computed between an inner ($r_m$) and an outer ($r_M$) radius.
If the galaxy surface brightness in the waveband $j$ is described by the Sersic
law
\begin{equation}
 \mu_j(r) = \mu_{0,j} + 1.086 \cdot b_j (r/r_{e,j})^{1/n_j} \ \ ,
 \label{SERSIC}
\end{equation}
we obtain
 \begin{eqnarray}
  \begin{array}{ll}
  \Delta(\mu_2 - \mu_1) / \Delta(\log r) = &  
  2.5 \log(e) \cdot \left[ \log(r_M) - \log(r_m) \right]^{-1} \cdot \\
   & \hspace{-2.5cm} \cdot \left[
  b_1 \cdot \left(  ( r_m / r_{e,1} )^{1/n_1} -
  ( r_M / r_{e,1} )^{1/n_1}  \right) + 
  b_2 \cdot \left(  ( r_M / r_{e,2} )^{1/n_2} - 
( r_m / r_{e,2} )^{1/n_2}  \right)   
\right]  \ ,
\end{array}
\label{GMACUMB}
\end{eqnarray}
\noindent
where $b$ is a function of $n$ ($b \sim 2n-1/3$, see
Caon, Capaccioli, \& D'Onofrio 1993) and the indices $1, 2$ denote the
two wavebands. In order to make a direct comparison with 
previous works \citep{PVJ90,SMG00} we set $r_m=0.1 \cdot r_{e,R}$ and
$r_M=r_{e,R}$.
Equation~\ref{GMACUMB} is more suitable for the present analysis than the
approximation of \citet{SPJ93}, in that it allows to address the case
of different values of $n$ and it is exact for arbitrary ratios of
the half-light radii.
We also verified that this formula gives a very close
approximation to the linear fit to the color profiles.

The $V-$, $R-$ and $K-$band structural parameters 
derived in the previous sections were
substituted in equation~\ref{GMACUMB} to derive the implied color gradients.
The uncertainties on the gradients were derived from propagation of 
the errors on $r_e$ and $n$, taking into account the covariance 
term.
The $V\!-\!R$ and $R\!-\!K$ color gradients derived 
by means of equation~\ref{GMACUMB} are 
shown in Figure~\ref{grad} as a function of $n$, and the relative 
median values are given in Table~\ref{tab_grad}.
The average optical ($V\!-\!R$) color gradient is  $-0.06^{+0.03}_{-0.15}$ 
mag/dex, in agreement with the $V-I$ gradients derived by \citet{TAO00}
at $z=$0.37 and with
the optical ($V,R,I$) gradients by \citet{SMG00} at $z\sim0.38$
for cluster early-type galaxies (see Figure 6 in \citet{SMG00} and Tables
2 and 3 in Tamura \& Ohta 2000).

The $V\!-\!R$ gradients present a weak trend with $n$:
galaxies with $n<4$ have larger negative gradients, with an average 
of -0.19 mag/dex, while for $n>4$ the average gradient is consistent
with zero or slightly positive (+0.07 mag/dex).
The wide majority of the galaxies have negative 
$R\!-\!K$ color gradients: the median value for the whole sample is 
$-0.49^{-0.43}_{-0.55}$ mag/dex. The trend with $n$ is
in this case stronger: the average $R\!-\!K$ color gradient is
$-0.73$ mag/dex for $n < 4$ and $-0.35$ mag/dex for 
$ n > 4$. 
Furthermore, the results of \S~6.1 indicate that  the 
color gradients are mainly due to the difference in $n$ 
for $n<4$ and to the difference in $r_e$ for $n>4$. 
The optical-NIR color gradients are thus related to the relevance 
of the disk component, as it is apparent from Figure~\ref{grad}: galaxies with
lower bulge-to-disk ratio ($n \rightarrow 1$)  have large $R\!-\!K$ color 
gradients, while bulge-dominated galaxies have moderate gradients.

In order to probe the origin of color gradients with the large wavelength baseline 
provided by the $K-$band, we adopted two different models in which color 
gradients are driven by (A) `pure age' or (M) `pure metallicity' gradients. 
We used synthetic colors from the code of \citet{BRC93} 
to simulate early--type galaxy spectra. 
An exponential star formation rate with time scale of 1 Gyr was adopted. 
In model (A) we assumed that the inner stellar population ($r \sim
r_m$) is 6.25 Gyr old at $z=0.31$. 
This value corresponds to the mean 
luminosity-weighted age of stellar populations in AC\,118 galaxies as 
inferred from the color -- magnitude relation (see Merluzzi et al. 2001). 
The stellar population age at the outer radius ($r \sim r_M$) was adjusted
to match the observed color gradients, and the metallicity was fixed 
to $Z=Z_{\odot}$.
In model (M) the inner stellar population is characterized by a metallicity 
$Z=1.8~Z_{\odot}$ \citep{TKA00}
and the observed color gradients were used to derive the 
outer $Z$ value. Age was fixed to 6.25 Gyr.
From the median values of the $R\!-\!K$ color gradients 
we derived the outer age and metallicity ($\tau_M$ and $Z_M$) 
for models (A) and (M) respectively. 
Such values were then used to predict the $V\!-\!R$ color gradients at 
$z=0.31$ and the $V\!-\!R$ and $R\!-\!K$ gradients at $z\sim 0$. 

Table~\ref{tab_grad_th} gives the predicted $V\!-\!R$ and $R\!-\!K$ gradients 
along with the gradients observed both at $z\sim0.3$ and at $z\sim0$.
Optical-NIR color gradients were measured for 12 nearby bright elliptical 
galaxies by Peletier et al. (1990a, b). 
From their data we obtain an average $R\!-\!K$ gradient
of $-0.14$, with a broad distribution of values ranging from $-0.62$ to $+0.15$.
\citet{PCD98} estimated an average $V\!-\!K$ gradient of -0.18 by applying
the formula of \citet{SPJ93} to the ratio of the relative
half-light radii, while an average $V-H$ gradient of $\sim-0.18$ can
be inferred in the same way from the Coma cluster $H$ and $V$
effective radii given by \citet{SGB98}.

The results shown in Table~\ref{tab_grad_th} may be summarized as follows.
The pure age model is able to explain the evolution of the $R\!-\!K$
color gradient between $z\sim0.3$ and $z\sim0$, but it fails to
describe the $V\!-\!R$ gradient at $z\sim0.3$. Conversely, the pure
metallicity model is able to reconcile the $V\!-\!R$ and $R\!-\!K$ color
gradients at $z\sim0.3$, but it fails to describe the evolution of the
$R\!-\!K$ gradient as a function of look-back time.

\section{Discussion and Conclusions}\label{CONC}

In this study we derived optical and NIR structural parameters for
galaxies belonging to the cluster AC\,118 at $z=0.31$ from ground-based
photometry. 
The present data set constitutes the first large sample of
structural parameters determined at intermediate redshifts in different
wavebands and, in particular, it is the first such sample in the near-infrared.
The samples consist in $N$=93 objects studied in the $R$ and $K$ wavebands 
and in $N$=58 galaxies studied in the $V-$band.

The reliability of the data was verified by comparison with the results 
from HST photometry for a subsample of galaxies: the scatter between 
measurements from HST- and from ground-based photometry is equivalent to the 
scatter between parameters determined with two different fitting methods 
from HST images, namely of $\sim$35\% in $r_e$. Given that these values 
of the scatter 
are generally claimed to be a measure of the `intrinsic uncertainty' in 
structural parameter measurements, we can state that 
reliable structural parameters can be still derived from ground-based
photometry in ordinary observing conditions at least up to $z\sim0.3$. 

Since morphology was not a selection criterion in this study, some remarks are 
needed for the comparison of our results with other works.
\citet{CBS98} established that $78\%$ of the galaxies in the core of AC\,118 
are early-types while $22\%$ are spirals. They also found a significant rise 
in the fraction of S0 galaxies with respect to the local 
density-morphology relation.
Morphological classification from \citet{CBS98} is available for 21 galaxies 
in sample D.
The number of spirals is 5, of which 3 have $n<$1.5 
while the other two have $n \gtrsim 4$. 
Only one of the 
spirals has `late-type' colors, while the others lie on the red sequence 
of AC\,118. It is interesting to notice that all the remaining galaxies with 
`late-type' colors lie outside of the cluster core and have $n < 2$. 
They are likely disk galaxies dominated by young stellar populations.
These facts are in agreement with a scenario in which spirals are accreted into 
the cluster from the field, cease to form stars, become gas-deficient objects 
and eventually undergo a morphological transformation \citep{CBS98}. 
The remaining galaxies with morphological information consist in 10 S0,  
eight of which with $2\lesssim n \lesssim 3$ and two with $n \sim 8$, and 
6 ellipticals, for which we find $n \gtrsim 4$. 
We stress that the galaxies with large disks and late-type colors are 
only $4\%$ of galaxies in sample D.

The main feature concerning the multi-band properties of galaxies is that 
these are much more concentrated in the NIR than in the optical. The mean ratio 
of NIR to optical half-light radii ($\sim 0.75$) is not different from that 
found in local samples of early-type galaxies ($\sim 0.8$, e.g. Pahre 1999), 
but the situation is complicated by the systematic differences found in the shape 
of the brightness distribution: galaxies with low $n$ value have a light profile 
more peaked in the center in the $K$-band ($n_K - n_R \sim 1$), while for 
$n \gtrsim 4$ a lower $r_{e,K}/r_{e,R}$ is observed 
($\sim 0.6$). Such trends can be interpreted as due to the increase of the 
disk fraction when $n$ approaches unity.  

To discuss the combined effects of half-light radii and Sersic indices, we 
derived the optical and NIR color gradients by means of equation 4.
Most of galaxies in AC\,118 do not show significant optical color gradients, with
a median value of $-0.06^{+0.03}_{-0.15}$. This result is consistent with
previous estimates of optical color gradients at intermediate redshifts 
\citep{SMG00,TAO00}. Conversely, strong negative gradients are found  
in optical--NIR colors: the median of the $R\!-\!K$ gradients amounts
to $-0.49^{-0.43}_{-0.55}$. Moreover, this effect depends on the shape of the light
profile: the 
$R\!-\!K$ gradient increases from $-0.73^{-0.63}_{-0.83}$ for 
galaxies with low $n$ value to $-0.35^{-0.29}_{-0.42}$ for $n \gtrsim 4$.    
Again this result can be explained by an increase in the disk fraction when 
$n \rightarrow 1$.

To interpret these results in terms of the properties of the stellar populations, 
we introduced a pure age and a pure metallicity model. We used the $R\!-\!K$
gradients to calibrate both models and to predict $V\!-\!R$ color gradients
at $z=0.31$. Moreover, the evolution of color gradients as inferred by the models 
was compared with the measurements by \citet{PDD90,PVJ90} for a sample of nearby 
elliptical galaxies. 
The age model is not able to reconcile our optical and NIR estimates at $z=0.31$   
since a significant negative gradient is predicted in $V\!-\!R$ ($\sim -0.26$).
On the other hand, the predicted color gradient agree with the measurements at 
$z=0$. Conversely, the pure metallicity model gives correct results at $z=0.31$ 
but predicts too strong NIR gradients at $z=0$. Since the age model is able to 
describe 
the evolution of color gradients, whereas the metallicity model correctly explains
the dependence of color gradients on wavelength, we argue that a model in which 
effects of age and metallicity are consistently combined could explain optical and 
NIR color gradients from $z=0.31$ up to $z=0$.
Finally, we remark that dust likely affects the optical-NIR color gradients.
Dust is actually presents in the core of most  
early-type galaxies \citep{GDJ95,VDF96} and is known to produce high central color 
gradients in bulges of S0 and spiral galaxies (e.g. Peletier et al. 1999).  

The distributions of both the optical and NIR half-light radii of galaxies 
at $z\sim0.3$ are found to be consistent with those of Coma early-type
galaxies. 
The result for the NIR waveband is the most significant,
since the NIR probes directly the mass distribution inside galaxies.
The observed distributions thus indicate that the sizes of bright galaxies did not 
substantially change since $z\sim 0.3$ to the present epoch. This result
suggests that (dissipation-less) merging did not play 
a major role in galaxy evolution during the last $4.4$ Gyr 
(see also Nelson et al. 2001).

\acknowledgments

We are grateful to Adam Stanford who provided us with his own photometry 
of AC\,118, that was used to check the calibration of our $R$-band data.
We thank the referee for his/her helpful comments and suggestions and 
Dr. M. D'Onofrio helpful discussions.
The observations at ESO were collected during the guaranteed time of 
the Osservatorio Astronomico di Capodimonte. 
This work is partially based on observations made 
with the NASA/ESA Hubble Space Telescope, obtained from the data archive at
the Space Telescope Science Institute. STScI is operated by the Association 
of Universities for Research in Astronomy, Inc. under NASA contract 
NAS 5-26555. Michele Massarotti is partly supported by a MURST-COFIN grant. 

\newpage

\appendix

\section{Calibration of HST data}
   \label{app_hstcalib}

AC\,118 was observed on November 1994 with the HST Wide-Field 
Planetary Camera 2 in the F702W filter. 
Two exposures of 2300 s and one exposure of 1900 s for
a total integration time of 6500 s are available for a single pointing. 

The HST images were calibrated into the Johnson $R-$band by means of the NTT
photometry. We proceeded  as follows.
The frames were smoothed to the same resolution of ground-based images 
by a convolution with the NTT $R-$band PSF. Aperture photometry was then performed with the 
SExtractor software \citep{BER96} in the same aperture of the ground data. 
Figure~\ref{HSTCALIB} shows the difference of $R$ and F702W instrumental 
magnitudes versus $V\!-\!R$ colors.
The data points were fitted with the equation:
\begin{displaymath}
 R = F702W_{instr} + a * ( V - R ) + b \label{COLEQ} \ \ .
\end{displaymath}
We found $a=30.93 \pm 0.10$ and $b=0.30 \pm 0.12$, and that the root mean square of 
the residuals to the fitted relation varies from 0.05 mag for $R\leq 20.5$ mag to 
0.1 mag for the whole sample. The few galaxies (9\%) which deviate more 
than $\sim0.2$ mag from the fit, turned out to be objects contaminated by 
bright neighbors. 
In the same figure we also plot the synthetic transformation to the $R$ waveband
of Holtzman et al. (1995, see their equation 9 and Table 10):
\begin{eqnarray}
 R & = & F702W_{instr}  + 0.486 * ( V - R ) - 0.079 * ( V - R )^2 + \nonumber \\  
 & + & 2.5 * \log (t_{exp}) + 2.5 * \log (GR) + \Delta m_{ap} + 
\Delta m_{exp} \label{SYNEQ} \ \ ,
\end{eqnarray}
where $t_{exp} $ is the exposure time,  $GR$ is the 
gain ratio of the WFPC2 chips, $\Delta m_{ap}$ is an aperture correction term derived
from the encircled energy measurements of \citet{HOL95}, and 
$\Delta m_{exp}$ is the long exposure correction term from Hill et al. (1998).
The same gain ratio $GR=2$ was adopted for each chip, since the effects
of gain differences are very small ($\lesssim 0.01$ mag) when compared to 
the other sources of uncertainty. We adopted $\Delta m_{ap}=+0.10$ mag and
$\Delta m_{exp}=+0.05$ mag. A further shift of $+0.05$ mag was needed to better 
match equation~\ref{SYNEQ} to our data. 
This is in agreement with previous results  of calibration of HST images 
with ground-based photometry (see J\o rgensen et al. 1999).

\section{PSF of ground-based images}  \label{app_gpsf}

The PSF of NTT ground-based images was described by the following model:
\begin{eqnarray}
 S(x,y) = \left[ G_1 + G_2 \right] \! \ast \! R \,  (x,y) \ \ ,
\label{PSF_EQ}
\end{eqnarray}
where $S$ is the PSF brightness distribution, $G_j (j=1,2)$ are two-dimensional
Gaussian functions and $R$ is a box function equal to 1 on the pixel area
and 0 elsewhere. The convolution in equation~\ref{PSF_EQ} was computed
from the one-dimensional convolution of a Gaussian with a rectangular function.
For the one-dimensional case, the following expression is obtained:
\begin{eqnarray}
 g \! \ast \! r \, (x) = (2 a)^{-1} & \!
  \cdot \left\{  Erf \left[ \frac{a/2+x}{ \sqrt{2} \sigma } \right] +
   Erf \left[ \frac{a/2-x}{ \sqrt{2} \sigma } \right] \right\} \ \ ,
\label{GAUSREC}
\end{eqnarray}
where g(x) is is a normal Gaussian function of width parameter $\sigma$, 
$r(x)$ is a rectangle of width $a$, and $Erf(x)$ is the error function.  
The analytical expression of the ground-based PSF is obtained by substituting 
the formula~\ref{GAUSREC} in equation~\ref{PSF_EQ}.

\section{Uncertainties on galaxy structural parameters} \label{app_err}

Galaxy structural parameters were derived by minimizing the right side of 
equation~\ref{CHIEQ}, i.e. by solving the following system of equations:
\begin{eqnarray}
 \frac{ \partial }{ \partial p_k} 
 \chi^2 (\left\{p_k\right\}; \left\{I_j\right\}, \left\{ s_l \right\}) = 0  \ \ , 
\label{NLSIST}
\end{eqnarray}  
where $\chi^2$ is a function of the fitting parameters $\left\{p_k\right\}$,
$\left\{I_j\right\}$ are the brightness values of image pixels, and 
$\left\{ s_l \right\}$ are the $PSF$ parameters for the ground-based data.
The changes  $\left\{\delta I_j\right\}$, $\left\{\delta s_l\right\}$ of 
$\left\{I_j\right\}$ and $\left\{ s_l \right\}$ determine variations 
$\left\{\delta \bar{p}_k\right\}$ of the solutions $\left\{\bar{p}_k\right\}$ 
of equations~\ref{NLSIST}. At the first order, $\left\{\delta \bar{p}_k\right\}$
are obtained from the following formulae:
\begin{eqnarray}
\delta \bar{p} = H^{-1} \cdot \left( A \cdot \delta I + B \cdot 
\delta s \right) \ \ ,
\label{ERR}
\end{eqnarray}  
where the symbol $\cdot$ denotes matrix multiplication, $\delta \bar{p}$, $\delta I$ 
and $\delta s$ are column vectors of components $\delta \bar{p}_k$, $\delta I_j$ 
and $\delta s_l$ respectively, $H^{-1}$ is the inverse of the $\chi^2$ Hessian matrix, 
$A$ and $B$ are rectangular matrices of components 
$\partial \chi^2 / \partial p_k \partial s_l$ and 
$\partial \chi^2 / \partial p_k \partial I_j$ 
respectively. All quantities are computed at $\left\{ \bar{p}_k\right\}$,
$\left\{ I_j\right\}$ and $\left\{ s_l\right\}$. \\
The noise of image pixels and the uncertainties on PSF parameters 
(see \S~\ref{PSFSEC}) were used to estimate the quantities 
$\left\{\delta I_j\right\}$ and $\left\{\delta s_l\right\}$. Uncertainties on 
fitting parameters were then derived by equations~\ref{ERR}.


\begin{figure}
\epsscale{0.6}  
\caption{ Dependence of the ground-based PSF on the chip position. 
The quantities shown in the figure are:
the ratio of the central intensities of the `outer' and `inner' Gaussians 
$I_2 / I_1$, the FWHM in the x- and y-  directions of the inner Gaussian 
($FWHM^{(1)}_x$ and  $FWHM^{(1)}_y$), the FWHM in the x- and y-  directions 
of the outer Gaussian ($FWHM^{(2)}_x$ and  $FWHM^{(2)}_y$), and the
asymmetry $A= FWHM^{(1)}_x / FWHM^{(1)}_y$ of the inner Gaussian. On the 
x-axis the distance to the frame centers $d$ is plotted. 
For each quantity, three panels are shown for the $V-$, $R-$ and $K-$bands 
as indicated in the upper-left plots.
The error bars correspond to $1\sigma$ standard intervals and the solid
lines are the curves used to create the PSF models for each galaxy
position (see text). 
}
\label{PSFFIT}
\end{figure}


\begin{figure}  
\epsscale{0.7}  
\caption{Results of the numerical simulations for $R$-band photometry.
The differences between input and output values are shown for the following
quantities: logarithm of $r_e$, $\langle\mu\rangle_e$, $m_T$, 
fundamental plane parameter $FP$, logarithm of the axis ratio $b/a$, and 
$n$. For each quantity, the two panels correspond to 
simulations with $r_e=0.3''$ and $r_e=0.6''$.
The total magnitude of the models is given on the abscissae. 
}
\label{SIM_MTOT}
\end{figure}


\begin{figure}  
\epsscale{0.7}  
\caption{Minimum $r_e$ for the applicability of the 2-D 
fitting to the ground-based data. The same quantities as in 
Figure~\ref{SIM_MTOT} are
plotted on the vertical axis while the logarithm of the 
`true' effective radius is shown on the abscissae.
For each quantity, the three panels correspond to the different values of
$n$ (see upper-left panels).
}
\label{SIM_RE}
\end{figure}


\begin{figure}
\epsscale{1.0}  
\caption{Two-dimensional fit of galaxies in the HST image by means of
Sersic and of $B+D$ models. 
Left and right pictures  are relative to galaxies with $F=+1$  and $F=0$
(see text). Upper panels: original image; middle left 
and right panels: Sersic and $B+D$ fitting models;
lower panels: corresponding residuals.
}
\label{BD_S}
\end{figure}


\begin{figure}  
\epsscale{0.7}
\caption{Distribution of the differences between structural 
parameters derived from the Sersic fit 
and from the $B+D$ decomposition for HST galaxies. 
From left to right, and from top to bottom: $r_e$, 
$\langle\mu\rangle_e$, $m_T$ and $FP$ parameter. 
Solid, dashed and dot-dashed lines are relative to galaxies with different 
fitting flag $F$ as indicated in the upper-left panel. Differences are in the
direction Sersic-($B+D$). 
}
\label{BD_S_PAR}
\end{figure}


\begin{figure}  
\epsscale{0.8}
\caption{Two-dimensional fit of galaxies in the HST F702W and NTT $R-$band 
images. For each quadrant, the left and right panels correspond to NTT 
and HST data respectively, and, from top to bottom, galaxy image, 
fitting model, and image of residuals are shown. 
HST and NTT images are scaled to the same counts/arcsec$^2$.
}
\label{HST_R}
\end{figure}


\begin{figure}  
\epsscale{0.8}
\caption{Comparison of structural parameters derived from NTT $R$ and 
HST F702W images. For each panel, the histograms show the 
fractional distribution of differences between NTT and HST parameters. 
The plotted quantities and the plotting ranges are the same as in 
Figure~\ref{BD_S_PAR}. Differences are in the direction NTT -- HST.
}
\label{conf_HST_R}
\end{figure}


\begin{figure}  
\epsscale{0.8}
\caption{Differences ($V-R$) of structural parameters derived 
from $V$ and $R$ images. 
The $R-$band $n$ is plotted on the abscissae. 
}
\label{V_R_PAR}
\end{figure}


\begin{figure}  
\epsscale{0.9}
\caption{ Distribution of optical (upper panel) and NIR (bottom) 
half-light radii of: 1) cluster galaxies at $z=0.31$ (solid line, this work);
2) cluster galaxies at $z=0.33$ (dashed line, Kelson et al. 2000b); early-type
galaxies in Coma ($z=0.023$, dash-dotted line, Pahre 1999).
The Coma radii were transformed to $z=0.31$ by adopting $\Omega_M=0.3$
and $\Omega_{\Lambda}=0$.
}
\label{Phare}
\end{figure}

\begin{figure}  
\epsscale{0.9}
\caption{Differences ($K-R$) of structural parameters derived from $R$ and $K$ 
images. The $R-$band $n$ is plotted in the abscissae. 
The plotting ranges are the same as in Figure~\ref{V_R_PAR}.
}
\label{R_K_PAR}
\end{figure}


\begin{figure}  
\epsscale{0.8}
\caption{ Color gradients of AC\,118 galaxies as derived from equation 4 
(see text). 
$V\!-\!R$ and $R\!-\!K$ color gradients are shown in the upper and lower 
panel respectively. The Sersic index in the $R-$band is plotted on the
abscissae. Error bars define $1 \sigma$ standard intervals.
}
\label{grad}
\end{figure}


\begin{figure}
\epsscale{1.0}
\caption{Calibration of HST data by ground-based NTT photometry.
Differences of $R$ and F702W instrumental magnitudes are plotted 
versus $V\!-\!R$. 
The solid line is the least squares fit to the data. 
The dashed curve represents the synthetic 
transformation of \citet{HOL95}.}
\label{HSTCALIB}
\end{figure}

\clearpage

\begin{deluxetable}{cccccc}
\tabletypesize{\scriptsize}
\tablecaption{Summary of the photometric data. \label{tbl-1}}
\tablewidth{0pt}
\tablehead{
\colhead{Band} & \colhead{Instrument}   & \colhead{Exp. Time}   &
\colhead{Field} &
\colhead{pixel scale}  & \colhead{seeing}     \\
& & (sec) & & (arcsec/pxl) & (arcsec) \\ 

}
\startdata
V & NTT-EMMI & 2700 & 9.$^\prime$15$\times$8.$^\prime$6 & 0.267 & 1.2  \\
R & '' & 1800 & '' & '' & 1.0  \\
K & NTT-SOFI & 15900 & 5$^\prime\times$5$^\prime$ & 0.292 & 0.8  \\
F702W & HST-WFPC2 & 6500 & $5.8 \, \mathrm{arcmin}^2$ & 0.1  & \\
\enddata
\label{AC118_DATA}
\end{deluxetable}

\clearpage

\begin{deluxetable}{lcccc}
\tabletypesize{\scriptsize}
\tablecaption{Description of the samples. \label{tab_SEL}}
\tablewidth{0pt}
\tablehead{
 & A & B & C & D \\ 
}
\startdata
 & & & & \\
 wavebands  & $\mathrm{F702W}$ & $R$,~$\mathrm{F702W}$ & $V$,~$R$ & $R$,~$K$ \\
 & & & & \\
 phot. redshift  & $all$ & $all$ & $0.3$ & $0.3$ \\
 & & & & \\
 selection   & $R \le 21$,~$r_e>0.15''$ & $R \le 21$,~$r_e>0.15''$ & $R \le 21$, $V \le 20.5$,~$r_e>0.15''$ & $R \le 21$,~$r_e>0.15''$ \\
 & & & & \\
 N & 50 & 37 & 58 & 93 \\
 & & & & \\
 fitting models & $B+D$,~$r^{1/n}$ & $r^{1/n}$ & $r^{1/n}$ & $r^{1/n}$ \\
 & & & & \\
\enddata 
\end{deluxetable}

\clearpage

\begin{deluxetable}{cccccccccccc}
\tabletypesize{\scriptsize}
\tablecaption{ Comparison of Sersic and $B+D$ fitting parameters. \label{tab_BD}}
\tablewidth{0pt}
\tablehead{
\colhead{F} & \multicolumn{2}{c}{$\Delta{\log r_e}$} & \colhead{} & \multicolumn{2}{c}{$\Delta{\langle\mu\rangle_e}$} & \colhead{} & \multicolumn{2}{c}{$\Delta{m_T}$} & \colhead{} &  \multicolumn{2}{c}{$\Delta{FP}$} \\ 
 \cline{2-3} \cline{5-6} \cline{8-9} \cline{11-12} \\
\colhead{ } & \colhead{$m$} & \colhead{$std$} & & \colhead{$m$} & \colhead{$std$} & & \colhead{$m$} & \colhead{$std$} & & \colhead{$m$} & \colhead{$std$}  \\
}
\startdata
    & & & & & & & & & & & \\
 -1 & $-0.06^{+0.09}_{-0.19}$  & $0.32^{0.47}_{0.29}$ & & $0.40^{+0.24}_{-0.90}$ &  $1.3^{1.8}_{1.2}$    & & $-0.10^{+0.26}_{-0.23}$ & $0.43^{0.95}_{0.37}$   & & $+0.06^{+0.10}_{+0.00}$    & $0.10^{0.14}_{0.09}$ \\
    & & & & & & & & & & & \\
  0 & $0.002^{+0.024}_{-0.031}$& $0.13^{0.16}_{0.12}$ & & $0.03^{+0.14}_{-0.10}$ & $0.56^{0.68}_{0.50}$  & & $0.04^{+0.07}_{-0.001}$ & $0.17^{0.21}_{0.15}$  & & $-0.011^{+0.001}_{-0.021}$    & $0.047^{0.057}_{0.042}$ \\
    & & & & & & & & & & & \\
  1 & $-0.06^{+0.003}_{-0.11}$ & $0.20^{0.30}_{0.17}$ & & $-0.20^{+0.07}_{-0.40}$ & $0.79^{1.20}_{0.66}$ & & $0.12^{+0.18}_{+0.04}$   & $0.24^{0.34}_{0.21}$   & & $-0.003^{+0.005}_{-0.02}$ & $0.036^{0.060}_{0.030}$ \\
    & & & & & & & & & & & \\
-1,0,1 & $-0.03^{+0.001}_{-0.056}$ & $0.185^{0.22}_{0.17}$ & & $-0.10^{+0.016}_{-0.21}$ & $0.76^{0.87}_{0.70}$ & & $0.04^{+0.08}_{+0.006}$ & $0.24^{0.28}_{0.22}$ & & $0.001^{+0.010}_{-0.007}$ & $0.06^{0.07}_{0.05}$ \\
    & & & & & & & & & & & \\
\enddata
\tablecomments{Column 1: fitting flag $F$. Columns 2, 3, 4 and 5: medians and standard 
deviations of the differences of the structural parameters. 
The upper and lower values define the $68\%$ confidence interval. Differences are in 
the direction Sersic - ($B+D$).} 
\end{deluxetable}

\clearpage

\begin{deluxetable}{lcc}
\tabletypesize{\scriptsize}
\tablecaption{ Comparison of HST and NTT $R-$band fitting parameters. 
\label{tab_HST_R}}
\tablewidth{0pt}
\tablehead{
\colhead{ } & \colhead{$m$} & \colhead{$std$} \\
}
\startdata
 & & \\
 $\Delta(\log r_e)$ & $-0.04^{+0.002}_{-0.08}$   & $0.16^{0.21}_{0.14}$ \\
 & & \\
 $\Delta(\langle\mu\rangle_e)$    & $-0.15^{-0.04}_{-0.26}$    & $0.67^{0.80}_{0.60}$ \\
 & & \\
 $\Delta(m_T)$      & $-0.08^{-0.03}_{-0.12}$    & $0.30^{0.33}_{0.28}$ \\
 & & \\
 $\Delta(FP)$       & $+0.020^{+0.037}_{+0.004}$ & $0.06^{0.08}_{0.05}$ \\
 & & \\
 $\Delta(\log b/a)$  & $+0.011^{+0.024}_{+0.003}$ & $0.082^{0.095}_{0.075}$ \\
 & & \\
 $\Delta(n)$        & $+0.01^{+0.45}_{-0.40}$ & $2.6^{2.9}_{2.4}$ \\
 & & \\
\enddata
\tablecomments{Median and standard deviation of quantities in column 1 are 
shown in columns 2 and 3. Upper and lower values define the $68\%$  
confidence interval. Differences are in the direction NTT--HST.}
\end{deluxetable}

\clearpage

\begin{deluxetable}{lcc}
\tabletypesize{\scriptsize}
\tablecaption{ Comparison of NTT $V-$ and $R-$band structural parameters.
\label{tab_V_R}} 
\tablewidth{0pt}
\tablehead{
\colhead{ } & \colhead{$m$} & \colhead{$std$} \\
}
\startdata
 & & \\
 $\Delta(\log r_e)$ & $+0.014^{+0.04}_{-0.004}$  & $0.17^{0.26}_{0.14}$ \\
 & & \\
 $\Delta(\langle\mu\rangle_e)$    & $+0.019^{+0.12}_{-0.07}$   & $0.75^{1.00}_{0.63}$ \\
 & & \\
 $\Delta(m_T)$      & $-0.045^{-0.016}_{-0.083}$ & $0.26^{0.34}_{0.22}$ \\
 & & \\
 $\Delta(FP)$       & $+0.008^{+0.020}_{-0.001}$ & $0.08^{0.09}_{0.07}$ \\
 & & \\
 $\Delta(\log b/a)$  & $-0.010^{-0.002}_{-0.017}$ & $0.06^{0.07}_{0.05}$ \\
 & & \\
 $\Delta(n)$        & $+0.024^{+0.27}_{-0.21}$   & $1.9^{2.2}_{1.8}$ \\
 & & \\
\enddata
\tablecomments{See Table~\ref{tab_HST_R}. Differences are in the direction $V-R$.}
\end{deluxetable}

\clearpage

\begin{deluxetable}{lcccccccccccc}
\tabletypesize{\scriptsize}
\tablecaption{ Comparison of NTT $K-$ and $R-$band structural parameters. 
\label{tab_R_K}}
\tablewidth{0pt}
\tablehead{
  & &  \multicolumn{3}{c}{$ all $} & \colhead{} & \multicolumn{3}{c}{$ n < 4 $} & \colhead{}  & \multicolumn{3}{c}{$ n > 4 $}  \\ 
  \cline{3-5} \cline{7-9}  \cline{11-13} \\
  & &   $m$ &  & $std$ & & $m$ &  & $std$ & & $m$ & & $std$ \\
}
\startdata
                    & & & & & & & & & & & & \\
 $\Delta(\log r_e)$ & & $-0.125^{-0.11}_{-0.14}$ & & $0.19^{0.21}_{0.18}$ & & $-0.09^{-0.07}_{-0.11}$ & & $0.12^{0.14}_{0.11}$ & & $-0.20^{-0.17}_{-0.24}$ & & $0.23^{0.26}_{0.21}$\\
                    & & & & & & & & & & & & \\
 $\Delta(\langle\mu\rangle_e)$    & & $-0.56^{-0.47}_{-0.63}$ & & $0.94^{1.02}_{0.89}$ & & $-0.37^{-0.29}_{-0.46}$ & & $0.68^{0.81}_{0.60}$ & & $-0.78^{-0.66}_{-0.93}$ & & $1.12^{1.23}_{1.05}$\\
                    & & & & & & & & & & & & \\
 $\Delta(m_T)$      & & $+0.04^{+0.09}_{-0.01}$ & & $0.26^{0.29}_{0.25}$ & & $-0.04^{-0.02}_{-0.10}$  & & $0.18^{0.22}_{0.16}$ & & $+0.15^{+0.23}_{+0.08}$ & & $0.29^{0.33}_{0.27}$\\
                    & & & & & & & & & & & & \\
 $\Delta(FP)$       & & $+0.04^{+0.05}_{+0.01}$ & & $0.11^{0.12}_{0.10}$ & & $+0.04^{+0.06}_{+0.015}$& & $0.09^{0.11}_{0.08}$ & & $+0.035^{+0.053}_{+0.015}$ & & $0.13^{0.14}_{0.12}$ \\
                    & & & & & & & & & & & & \\
 $\Delta(\log b/a)$  & & $-0.002^{+0.005}_{-0.009}$ & & $0.060^{0.067}_{0.055}$ & & $-0.005^{+0.007}_{-0.014}$ & & $0.060^{0.067}_{0.056}$ & & $+0.000^{+0.007}_{-0.008}$ & & $0.60^{0.76}_{0.52}$ \\
                    & & & & & & & & & & & & \\
 $\Delta(n)$        & & $+0.52^{+0.60}_{+0.48}$ & & $2.0^{2.2}_{1.9}$ & & $+0.95^{+1.0}_{+0.85}$ & & $1.6^{1.9}_{1.5}$ & & $+0.08^{+0.15}_{+0.02}$ & & $2.2^{2.5}_{2.0}$ \\
                    & & & & & & & & & & & & \\
\enddata
\tablecomments{Medians and standard deviations are shown for galaxies with
different ranges of $n$. Differences are in the direction $K-R$.}
\end{deluxetable}

\clearpage

\begin{deluxetable}{lccccccc}
\tabletypesize{\scriptsize}
\tablecaption{Median values of $V\!-\!R$ and $R\!-\!K$ color gradients of AC\,118 galaxies.
\label{tab_grad}}
\tablewidth{0pt}
\tablehead{
 & & $all$ & & $n < 4$ & & $n > 4$ \\
}
\startdata
 & & & & & & \\
 $(V-R)$ & & $-0.06^{+0.03}_{-0.15}$ & & $-0.19^{-0.09}_{-0.28}$ & & $+0.07^{+0.18}_{-0.04}$ \\
             & & & & & & \\
 $(R-K)$ & & $-0.49^{-0.43}_{-0.55}$ & & $-0.73^{-0.63}_{-0.83}$ & & $-0.35^{-0.29}_{-0.42}$ \\
 & & & & & & \\
\enddata
\tablecomments{ Color gradients are in units of mag/dex. }
\end{deluxetable}

\clearpage
\begin{deluxetable}{ccccc}
\tabletypesize{\scriptsize}
\tablecaption{Computed and observed color gradients. \label{tab_grad_th}}

\tablewidth{0pt}
\tablehead{
 MODEL                & $grad(V-R)$  & $grad(V-R)$ & $grad(R-K)$ & $grad(R-K)$ \\
                      &    $z\sim0.3$   &    $z\sim0$     &    $z\sim0.3$   &    $z\sim0$    \\
}
\startdata
 & & & \\
 AGE                  &    -0.26     &   -0.04     &     -0.49\tablenotemark{a}   &     -0.13   \\
 $\tau_M=3.2Gyr$      & & & &\\
                      & & & \\
 METALLICITY          &    -0.15     &   -0.10     &     -0.49\tablenotemark{a}    &     -0.41    \\
 $Z_M=0.5Z_{\odot}$     & & & &\\
                      & & & &\\
 observed             &  -0.1\tablenotemark{b} &   -0.02\tablenotemark{c}  &     -0.49\tablenotemark{d} &  -0.14\tablenotemark{c}    \\
                      & & & &\\
\enddata
\tablenotetext{a}{ Constrain to the model.}
\tablenotetext{b}{ See Tamura \& Ohta (2000); Tamura et al. (2000); Saglia et al. (2000), 
and this work. }
\tablenotetext{c}{ See Peletier et al. (1990a, b). }
\tablenotetext{d}{ This work. }

\tablecomments{The predictions are based on pure age 
and pure metallicity gradient models (see text). Column 1: adopted model. 
Column 2, 3: optical -- optical color gradient at $z\sim0.3$ and $z\sim0$ 
respectively. Column 4: optical -- NIR color gradient at $z=0$.
color gradients are reported in units of mag/dex.}
\end{deluxetable}

\end{document}